\documentclass[aps,prl,superscriptaddress,showpacs,floatfix,amsmath,amssymb,twocolumn]{revtex4}
\usepackage{epsfig}
\usepackage{bm}
\usepackage{hyperref}
\usepackage{balance}
\usepackage[usenames]{color}
\usepackage[normalem]{ ulem }
\begin{document}

\title{Reduced spin-orbit splitting in $^{35}$Si: weak binding or density-depletion effect ?
}
\author{O.~Sorlin} 
\affiliation{Grand Acc\'el\'erateur National d'Ions Lourds (GANIL),
CEA/DSM-CNRS/IN2P3, Bvd Henri Becquerel, 14076 Caen, France}
\author{F. de~Oliveira Santos} 
\affiliation{Grand Acc\'el\'erateur National d'Ions Lourds (GANIL),
CEA/DSM-CNRS/IN2P3, Bvd Henri Becquerel, 14076 Caen, France}
\author{J.P. Ebran} 
\affiliation{CEA, DAM, DIF, F-91297 Arpajon, France} 
\affiliation{Universit\'e Paris-Saclay, CEA, Laboratoire Mati\`ere en Conditions Extr\^emes, 91680, Bruy\`eres-le-Ch\^atel, France}

\begin{abstract}
The reduction of the neutron spin-orbit (SO) splitting $2p_{3/2} - 2p_{1/2}$ between the $^{41}$Ca and $^{35}$Si isotones is a unique feature throughout the chart of nuclides, as the SO splitting usually increases with decreasing $A$. The way it is reduced, either gradually between $^{41}$Ca and $^{35}$Si or abruptedly between $^{37}$S and $^{35}$Si, as well as the origin of its reduction, whether from the weak binding energy of the  $p$ states or from the sudden depletion in the central proton density moving to $^{35}$Si, are subject of debate. The results reported here using the self-consistent Covariant Energy Density Functional calculations with the DD-ME2 parametrization rather point to an abrupt, local decrease in $^{35}$Si, and to the large dominance of the central density depletion effect. It is concluded that  weak binding, central density depletion as well as correlations must be taken into account to fully evaluate the amplitude and causes of this SO splitting reduction. To broaden the scope of the present work, discussions also include the recent experimental results of proton and neutron SO splittings around $^{132}$Sn and $^{56}$Ni.
\end{abstract}

\maketitle

\begin{figure}[h]
\includegraphics[width=\columnwidth]{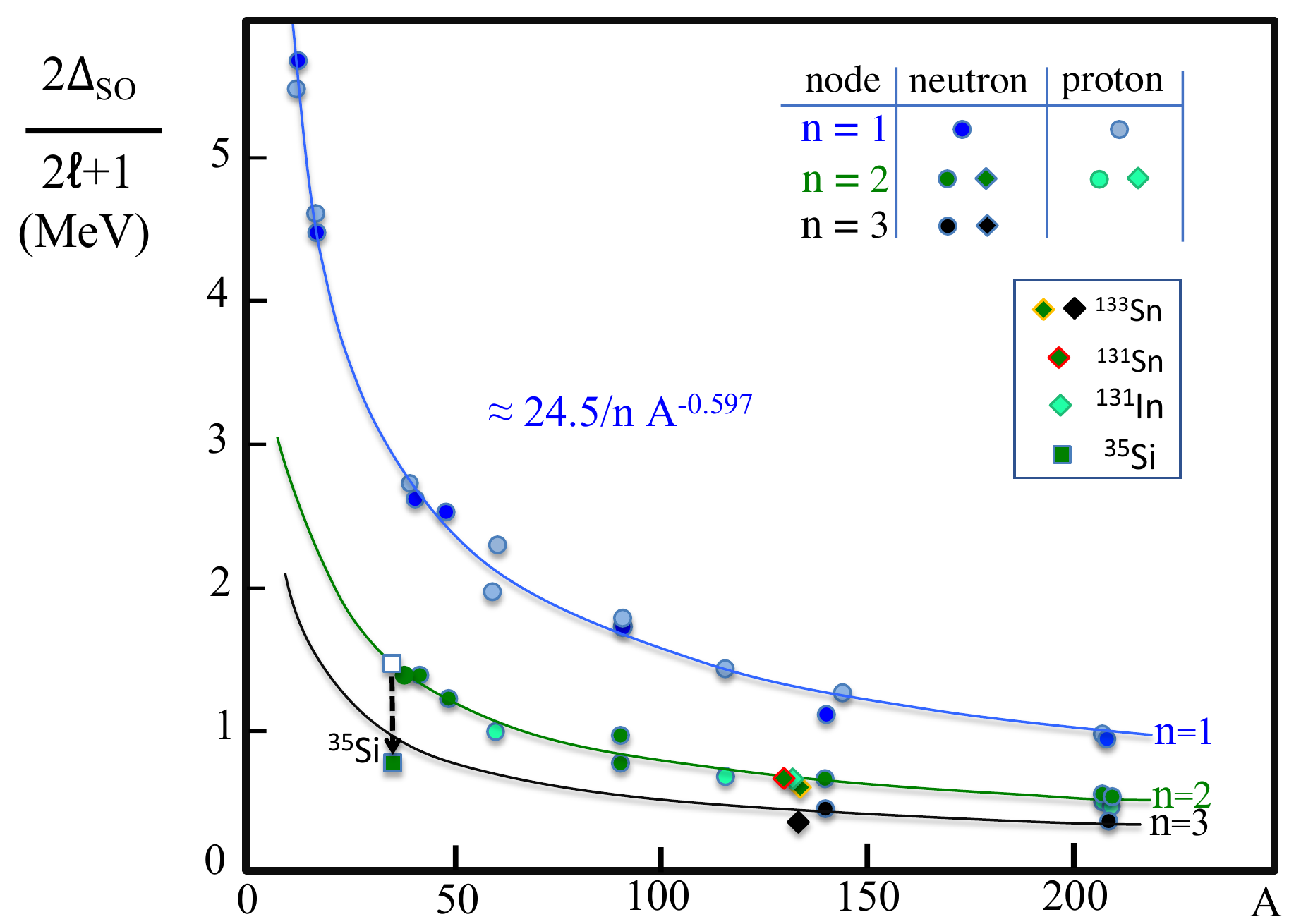}
\caption{Evolution of the SO splitting 2$\Delta_{SO}$/$(2\ell+1)$  as a function of the atomic mass A for different values of $n$, with $n$ corresponding to  the number of nodes of the nucleon's radial wave function. This representation, derived in Ref. \cite{Mair93} for the $p,d,f,g,h,i$ orbits with $n=1-3$,  is meant to show the global SO trend and not to provide exact values. Experimental results obtained from the states carrying the dominant single-particle or single-hole strength in $^{133}$Sn \cite{Jone10}, $^{131}$In \cite{Tapr14},  $^{131}$Sn \cite{Orla18},  $^{37}$S \cite{ENSDF} and $^{35}$Si \cite{Burg14} have been included. The $^{35}$Si data point strongly deviates from the expected trend.}
\label{SO-Mairle} 
\end{figure}

\noindent {\it Introduction- } The spin-orbit (SO) force, which results from the coupling of the particle orbital momentum $\ell$ and its intrinsic spin value $s$, plays an important role in quantum systems \cite{Ebra16}. In analogy with atomic physics, this force was introduced empirically in atomic nuclei in 1949 \cite{Gopp49} to account for shell gaps and magic numbers that could not be explained otherwise. In a simplified version, the corresponding SO potential can be written as $V_{SO}$= - $V_{\ell s}$ $ \frac{\partial \rho(r)}{ \partial r} $ $ \vec{\ell}  \cdot \vec{s}$, where $\rho(r)$ is the density distribution of the nucleons.  The scalar product lifts the degeneracy between the $j=\ell+s$ and $j=\ell-s$ spin-orbit partners, the latter being the least bound. The resulting SO splitting $\Delta_{SO}$ is of the order of the nucleon's binding energy and scales approximately with 24.5/n ($\ell$+1/2) A$^{-0.597}$ \cite {Mair93}, as shown in Fig. \ref{SO-Mairle}, with $n$ the number of nodes in the wave function and A the atomic mass of the nucleus. This relation just considers the bulk variation of the SO force and not local effects of two-body LS or tensor forces  that would further modify the SO splittings, when certain combinations of protons and neutrons orbitals are occupied \cite{Gaud06, Sorl08, Smir12, Otsu20}.

More realistic versions of this force can be found for mean field \cite{Bend99} and relativistic mean field theories \cite{Shar95,Lala98}. They usually differ by their isospin dependence \cite{Ebra16b} but their common feature is their density dependence. This implies an energy reduction between SO partners in weakly bound nuclei having a diffuse surface density-distribution (e.g. \cite{Shar95,Lala98}), as well as for orbits probing the interior of a nucleus with a central density depletion (e.g. \cite{Bend99, Todd04, Li16, Muts17,Dugu17}). Although shell-model calculations do not have explicit density-dependent two-body matrix elements, the lack of occupation of low-$\ell$ orbits also induces a central density depletion \cite{Grass09}. However, the correct long-tail  behavior of the loosely bound orbits' wave function cannot be properly treated in a Harmonic Oscillator basis, which is used in most of the shell model calculations \cite{Sign11,Yuan14}. The Gamow Shell Model approach overcome this limitation and allows the description of open quantum systems \cite{Mich05}. 

Nuclei exhibiting central density depletions are very rare (see e.g. \cite{Grass09}), but  $^{34}$Si likely has such a property. Ref. \cite{Muts17} has shown, using proton knockout reactions, that the $2s_{1/2}$ proton orbital is filled in $^{36}$S and almost empty in $^{34}$Si. As a large fraction of the radial part of this $\ell=0$ orbital is peaked in the center of the nucleus, the lack of $2s_{1/2}$ naturally induces a central density depletion. This is however not a direct proof  of a central depletion and electron scattering measurement would be ideal to ascertain this change in central density between $^{36}$S and $^{34}$Si. However, in the meantime, our hypothesis is supported by the $(e,e')$ experimental work of Cavedon et al. \cite{Cave82}, which has shown that the difference in charge density distributions between $^{206}$Pb and $^{205}$Pb is entirely reproduced, from the interior to the surface of the nucleus, by the removal of 0.7 protons from the $3s_{1/2}$ orbital.  In other words, the lack of $s_{1/2}$ occupancy directly translates into a reduction in central density, a feature that is also supported by many theoretical approaches (see, e.g. \cite{Grass09,Li16, Dugu17,Saxe19}).


The $3/2^-$ and $1/2^-$ states in $^{35}$Si, coming from the neutron $2p_{3/2}$ and $2p_{1/2}$ SO partners, fulfill these two conditions of being potentially sensitive to the weak binding effect and to the nuclear density depletion in $^{34}$Si. Indeed, while the $7/2^-$ ground state of $^{35}$Si is bound by 2.470(40) MeV, the $3/2^-$ and $1/2^-$ spin-orbit partners are bound by only 1.560(40) and 0.426 (40) MeV, respectively, according to Ref. \cite{Burg14}. Moreover, the proton density distribution of $^{35}$Si likely exhibits, as for $^{34}$Si, a significant central depletion due to the lack of $2s_{1/2}$ protons \cite{Muts17}. Consequently, the reduction of the $2p_{3/2}- 2p_{1/2}$  splitting, that is observed between $^{37}$S and  $^{35}$Si (see Fig. \ref{SO-Mairle}), can be caused by the two effects. The experimental and theoretical works of Refs. \cite{Todd04, Burg14,Li16, Muts17, Dugu17} point out that the central density depletion is the major cause of the SO splitting reduction, while the work of Kay {\it et al.} \cite{Kay17} finds that the reduction between spin-orbit partners can be fully explained by the effect of weak binding energy. Important is to say that the two-body tensor force plays no role between these two nuclei, as the protons involved lie in an $\ell$ = 0 ($2s_{1/2}$) orbital. 

In this Letter, we wish to further substantiate the role of these two competing effects. We shall first briefly summarize the work of Kay {\it et al.}, then present another experimental viewpoint to address the decrease in SO splitting in the $N=21$ isotones, before showing the results of two different theoretical approaches, each of them having more self-consistency, as compared to the work of Ref. \cite{Kay17}. We finally broaden the scope of the present study by including recent experimental works related to proton and neutron SO splitings around the doubly-magic $^{132}$Sn and $^{56}$Ni nuclei. 

\begin{figure}[h]
\includegraphics[width=\columnwidth]{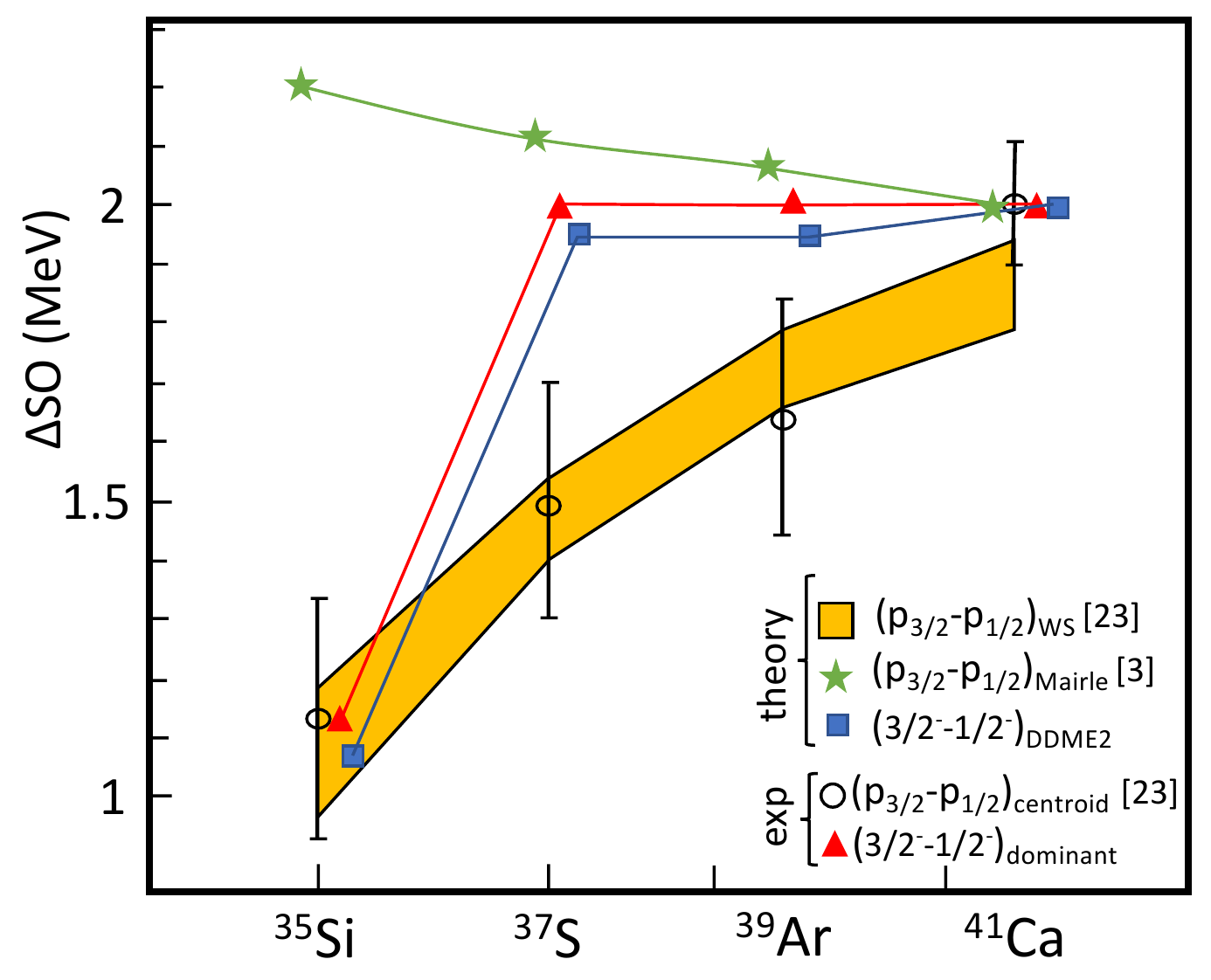}
\caption{Evolution of the $p_{3/2}$ - $p_{1/2}$ or 3/2$^-$ - 1/2$^-$ splitting, $\Delta SO$, for the $N=21$ isotones. Black open circles (with estimated error bars)  correspond to the centroid of the single-particle strength derived in \cite{Kay17}, in which WS calculations were made (orange band).  Green stars correspond to the systematics of Ref. \cite{Mair93}. Red filled triangles are obtained using the energy difference between the  $3/2^-$ and  $1/2^-$ states having  the dominating spectroscopic factor value, when populated by the $(d,p)$ reaction. Blue squares correspond to Covariant Energy Density Functional calculations with the DDME2 parametrization of the $3/2^-$ and  $1/2^-$ states shifted upwards by 340 keV (see text). Some symbols have been slightly shifted to the left or right to be better distinguished. }
\label{Delta-SO} 
\end{figure}

\noindent {\it  Reduced energy between SO pairs due to weak binding- }  Kay {\it et al.} \cite{Kay17} have determined a gradual decrease of the SO splitting from $^{41}$Ca$_{21}$ to $^{35}$Si$_{21}$, shown with black open circles in Fig. \ref {Delta-SO}, by collecting the single-particle and single-hole strength of 3/2$^-$ and 1/2$^-$ states available in all studied nuclei. They added  error bars to their experimental values to take into account some unknown unobserved strength. They have used a one-body Woods-Saxon (WS) potential with a fixed spin-obit amplitude $V_{\ell s}$ = 6 MeV to reproduce the $2p_{3/2}-2p_{1/2}$ splitting in $^{41}$Ca, and the WS potential depths were adjusted in each $N=21$ isotone to match the binding energy of the $2p_{3/2}$ orbital. They calculated the energy of the $2p_{1/2}$ and determined the corresponding SO splittings, shown in the orange band of Fig. \ref {Delta-SO}. From the good agreement between the experimental and calculated trends, they affirm that simple geometrical effects on loosely bound $p$ states can $\it fully$ account for a {\it gradual} reduction of the neutron $2p_{3/2}- 2p_{1/2}$ splitting between $^{41}$Ca and $^{35}$Si, {\it without} any need for a modification of the spin-orbit coupling. 

An important remark is that, by applying a constant SO splitting of about 2 MeV to the weakly bound $2p_{3/2}$ orbital at 1.560 MeV in $^{35}$Si, the $2p_{1/2}$ orbital becomes unbound by about 500 keV. In their calculation, this latter orbital is therefore extremely sensitive to the effect of coupling to continuum. Solving the Schr$\ddot{o}$dinger equation for the $2p_{1/2}$ orbital makes it bound, which results in a very significant apparent change of the $2p_{3/2}- 2p_{1/2}$ splitting. If, as will be discussed in the following, the SO splitting is reduced by the central density depletion, the $2p_{1/2}$ orbital would become more bound and less sensitive to the effect of weak binding.

\noindent{\it Evolution of SO splitting in the $N=21$ isotones-}  Kay {\it et al.} \cite{Kay17} collected the single-particle and single-hole strength available in all studied nuclei to deduce a decrease of the $2p_{3/2}- 2p_{1/2}$  SO splitting in the $N=21$ isotones.  However, the amplitude of this apparent reduction could not be firmly quantified from the determination of single-particle energies, as this information could never be obtained experimentally \cite{Dugu15}. Moreover, the available experimental  information on single-particle or single-hole strength strongly differs between the $N=21$ isotones. Many $3/2^-$ and $1/2^-$ states are observed in $^{41}$Ca, while fewer and fewer are observed when $Z$ decreases until reaching $^{35}$Si, in which only one $3/2^-$ and one $1/2^-$ states have been observed in \cite{Burg14}.  

By using the $p$ fragments (3/2$^-$ and 1/2$^-$) that are populated with the largest $(d,p)$ spectroscopic factor instead, the SO splitting $\Delta_{SO}$ of Fig. \ref {Delta-SO} (shown with red filled triangles) remains remarkably constant between $^{41}$Ca, $^{39}$Ar and $^{37}$S at a value around 2 MeV (with only 0.5\% variation), and decreases suddenly to 1.134 MeV in $^{35}$Si. This trend differs then completely from that of a gradual decrease. 

Being aware of the fact that these dominant fragments carry some amount of correlations, this evolution was compared  to shell model \cite{Burg14} and {\it ab initio} calculations  \cite{Dugu17}, in which correlations are taken into account. In particular, Fig. 21 of  Ref. \cite{Dugu17} shows a clear correlation between the increase of charge density depletion and the reduction of the SO splitting between $^{37}$Si and $^{35}$Si, meaning that without depletion, the SO splitting should not change. This is at variance with Ref. \cite{Kay17}, in which the reduction of SO splitting between $^{41}$Ca and $^{35}$Si is gradual and solely attributed to a progressive decrease of the neutron binding energy, independently of the occupied proton orbitals.

It is also very instructive to notice that the presently discussed change of the $p$ SO splitting with $A$, would it be gradual or sudden, is opposite to the global trend predicted by Mairle \cite{Mair93} (green stars in Fig. \ref {Delta-SO}), that has been derived from the study of SO splitting throughout the chart of nuclides (see Fig. \ref{SO-Mairle}). This significant discrepancy is extremely rare, but it suggests that physics exist beyond the global increase of SO splitting with decreasing $A$, likely in specific two-body forces and in effects due to the proximity of the continuum  \cite{Hamm01,Orla18}, for the least bound states. 

{\it Woods Saxon calculations -} The one-body Woods-Saxon (WS) potential approach has certainly a relatively good predictive power between mirror nuclei, once one of the nucleus has a well established structure (see, e.g. \cite{Fort95,Stef14}). In Ref. \cite{Stef14}, the profound reordering of the levels between $^{16}$N and the unbound $^{16}$F was successfully explained.  However, when changing nuclei, this predictive power is lost as this method is not self-consistent, i.e. the density distribution or the shape of the mean field potential does not change consistently as different orbitals are occupied, when moving from one nucleus to another. Abrupt changes in structure of the nuclei are however rare, making this simplified WS approach still used with a certain success \cite{Orla18}. Coming back to the $N=21$ isotones, a normalization of the depth of the WS potential had to be made in order to get the $2p_{3/2}$ orbit at the correct binding energy in \cite{Kay17}. Using this method, the evolution of the $2p_{3/2}$ binding energy, one of the member of the SO splitting, is therefore not predicted but fixed.

\begin{figure}[h]
\includegraphics[width=\columnwidth]{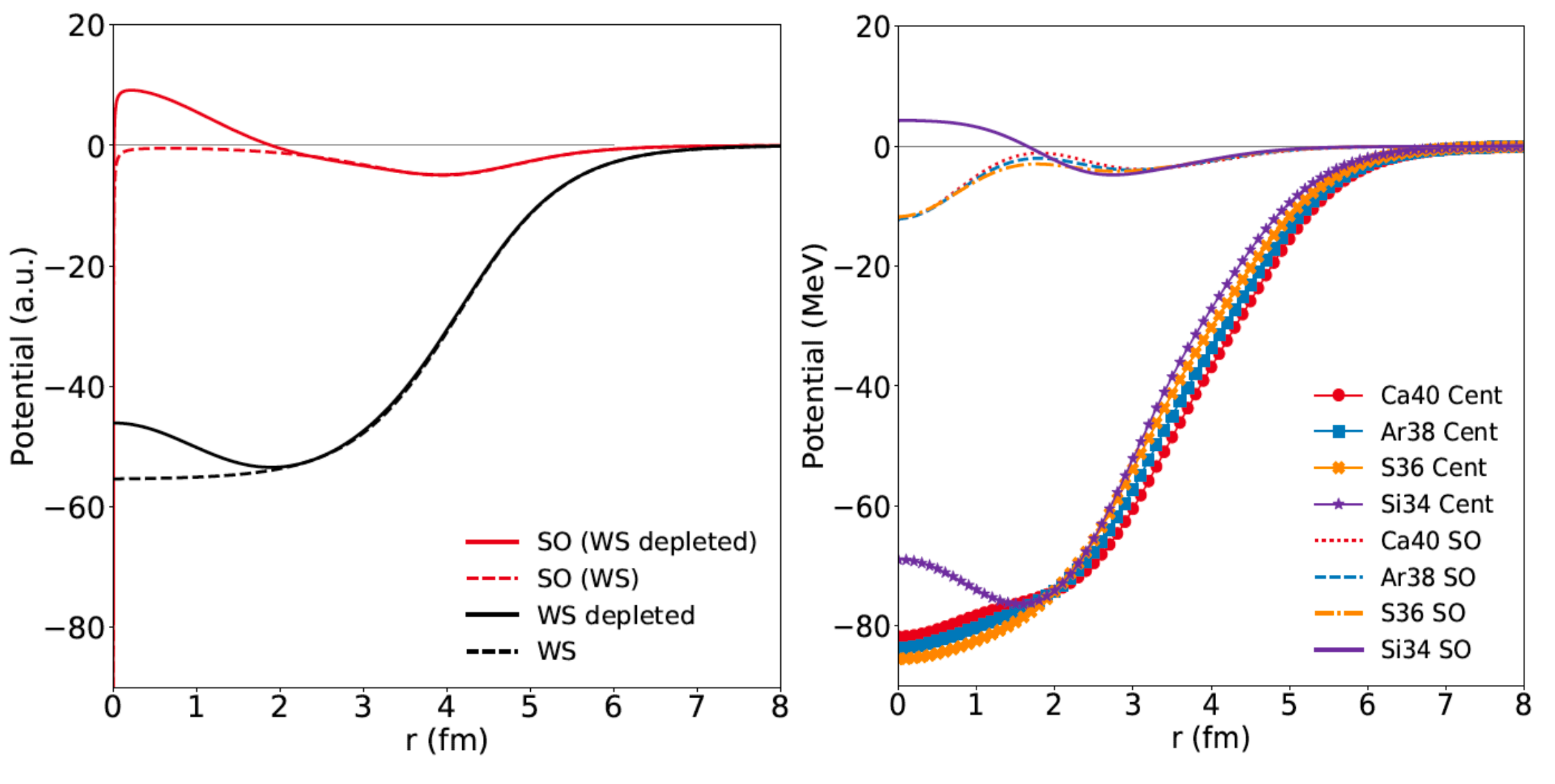}
\caption{Left: Woods-Saxon (WS)  (black dashed line) and SO potentials (red dashed line) are shown as a function of the radius in $^{34}$Si. The continuous lines show the effects induced by the removal of a pair of $2s_{1/2}$ protons. Right: Calculated DD-ME2 nuclear  and spin-orbit potentials for the $N=20$ isotones.  A qualitatively good agreement is found with the WS potential, except that the DD-ME2 calculation does not find a flat-bottom shape at low radius for the S-Ca nuclei, hence leading to a negative SO potential around the center of the nucleus rather than 0.}
\label{pot} 
\end{figure}

It is however possible to use this simple potential approach to access to the influence of the central depletion on the energy of the orbits.  We have performed calculations with the WS potential of Ref. \cite{Kay17} ($r_0=1.28$ fm, $a$ = 0.63~fm, $V_{\ell s}$ = 6~MeV), except that it was modified to take into account the existence of a central density depletion, due to the lack of two $2s_{1/2}$ protons. The shape of the depletion to be applied to the potential well was determined consistently from the calculation of the $2s_{1/2}$ wave function in the WS potential. The SO potential has been derived accordingly, as shown in the left part of Fig. \ref{pot}. The calculated $2p_{3/2}- 2p_{1/2}$ SO splitting is found to be of 1.8~MeV, under the assumption of no central depletion at an hypothetical binding energy of 6 MeV for the $2p_{3/2}$ orbit. This splitting is reduced to 1.35 MeV when applying the central density depletion of Fig. \ref{pot}. It further decreases to 0.91~MeV from the effect of the weak binding, when using the experimental binding energy of the $2p_{3/2}$ orbital in $^{35}$Si. In contrast with Ref. \cite{Kay17}, we find that adding a central depletion in $^{35}$Si, which corresponds to add some self-consistency to the WS approach,  the SO splitting  is reduced by about 50\% from the central depletion and 50\% from the weak binding. The effect of the central depletion in changing the $2p$ SO splitting is probably however underestimated as the WS potential has a  flat-bottom shape, contrary to the more realistic calculation (see left and right parts of Fig. \ref{pot}). With the use of this WS approach, we conclude that, when not considering the proton depletion into account, as in \cite{Kay17}, the effect of the continuum is clearly overestimated. 

{\it Covariant Energy Density Functional calculations- } In order to judge the combined effects of a central proton density depletion (induced by the removal of $2s_{1/2}$ protons) and of the weak neutron binding energy, in inducing a reduction of the $p$ SO splitting between $^{37}$Si and $^{35}$Si ($\Delta SO$), we use a Covariant Energy Density Functional (CEDF) calculation with the DD-ME2 parametrization \cite{DDME2}. The relativistic mean-field equations are solved in a large size box in order to obtain the correct asymptotic behavior of the orbitals. The calculated total and SO neutron potentials, to be compared with those obtained in the WS approach, are shown  in the right part of Fig. \ref{pot}. 


A first way to evaluate the $2p_{3/2}- 2p_{1/2}$ SO splitting in the $N=21$ isotones is via the calculation of the one-neutron removal energy $E(Z,N=21) - E(Z,N=20)$, where $E(Z,N)$ is the binding energy of the $(Z,N)$ nucleus. In the present CEDF calculations, the odd-N nuclei are computed within the so-called equal filling approximation~\cite{ber03,per08}. After calculating the removal energies for the $N=21$ systems in the 3/2$^-$ and 1/2$^-$ states, the $2p_{3/2}- 2p_{1/2}$ SO splittings, $E_{1/2^-}(Z,N=21) - E_{3/2^-}(Z,N=21)$, are given in Table~\ref{tab:dso}. Energies have globally been shifted upwards by 340 keV in Fig.~\ref{Delta-SO} to match the experimental value in $^{41}$Ca. 
\begin{table}[h]
\begin{tabular}{|c|c|c|}
\hline  
Isotope     &   $\Delta$SO odd (MeV)  & $\Delta$SO even (MeV)  \\
\hline  
Si          &  0.72                  &  0.76                   \\
\hline
S           &  1.51                  &  1.64                   \\
\hline
Ar          &  1.50                  &  1.67                   \\
\hline
Ca          &  1.66                  &  1.71                   \\
\hline
\end{tabular}

\caption{Magnitudes of $2p_{3/2}- 2p_{1/2}$ SO splittings evaluated by subtracting the binding energies 
of the $N=21$ nuclei in the 3/2$^-$ and 1/2$^-$ states ($\Delta$SO odd) or
by subtracting the energies of the $2p_{1/2}$ and the $2p_{1/2}$ orbitals
in the $N=20$ systems ($\Delta$SO even). See text for details.} 
\label{tab:dso} 
\end{table}

\begin{figure}[t]
\includegraphics[width=\columnwidth]{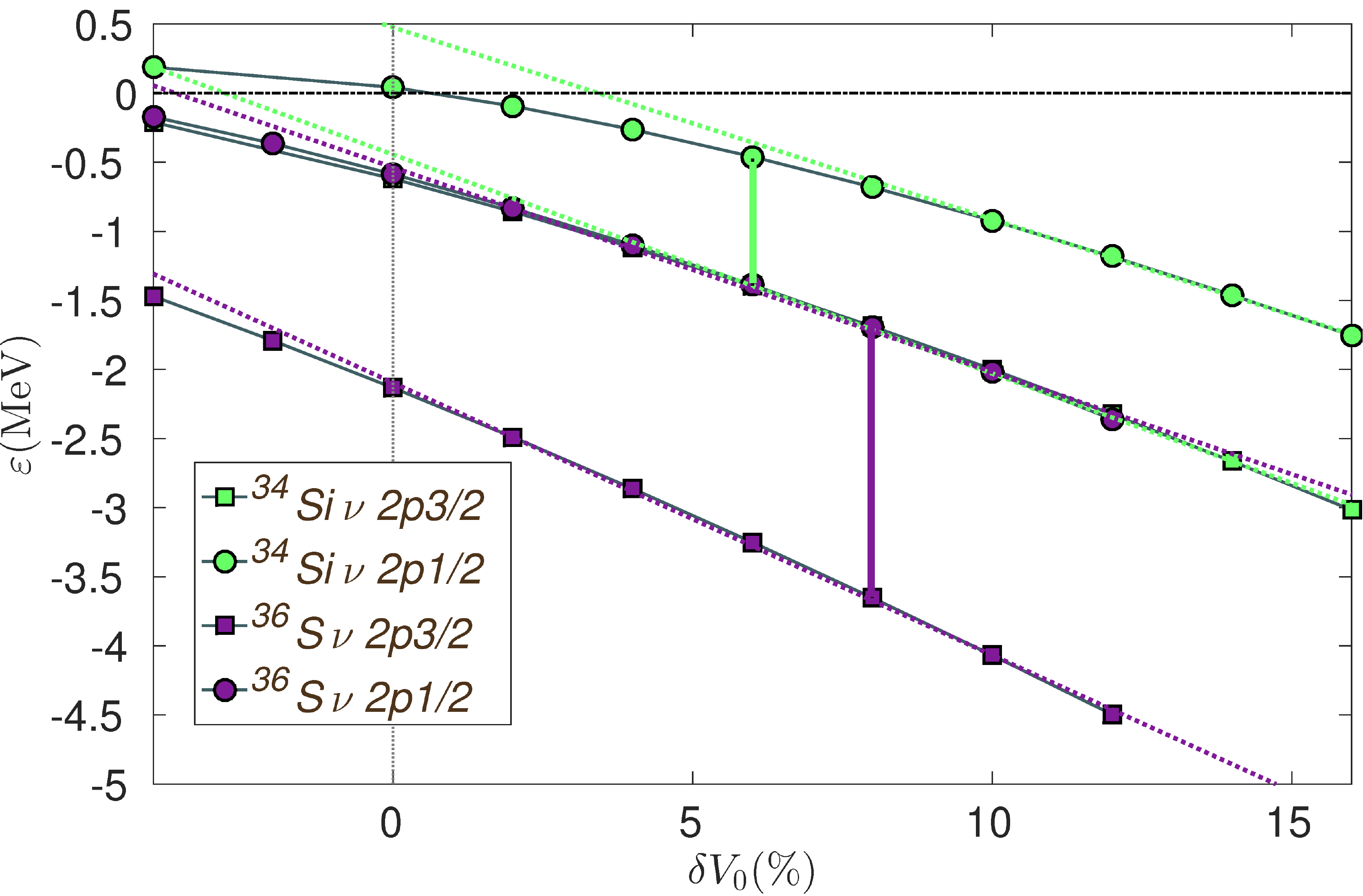}
\caption{Evolution of the neutron $2p_{3/2}$  (square symbols) and $2p_{1/2}$ (round  symbols) single-particle energies $\epsilon$ in $^{36}$S (violet) and $^{34}$Si (green) as a function of the relative change of the depth of the confining potential by $\delta V_0$ (in \%).  The nominal calculations of the SO splittings, indicated by the dashed line at $\delta V_0$ = 0, are also reported in the second column of Table \ref{tab:dso}. The violet and green vertical  lines indicate the amplitude of calculated SO splittings at the experimental binding energies of $^{37}$S and $^{35}$Si, respectively. }
\label{SO} 
\end{figure}

Another way of computing the evolution of $2p_{3/2}- 2p_{1/2}$ SO splitting is by subtracting the energies of the $2p_{1/2}$ and $2p_{3/2}$ orbitals in the even-even $N=20$ systems. Indeed, according to the  Koopmans' theorem~\cite{koo34}, the one-neutron removal energies of the odd nuclei are close to the Hartree-Fock single-particle energies calculated in the even-even system $(Z,N=20)$. As can be seen in (Table~\ref{tab:dso}), both methods yield similar results. The slight differences come from the fact that the Koopmans' theorem neglects the rearrangement of the neutron orbitals when going from the $N$ to the $N+1$ system. 

The second method is more convenient to track and quantify the impact of the weak neutron binding energy on the $2p_{3/2}- 2p_{1/2}$ SO splitting. To this end, we introduce an artificial change in the depth of the self-consistent confining potential $\delta V_0$, as compared to $\delta V_0$=0 used in Table  \ref{tab:dso}.  As shown in Fig. \ref{SO}, the major reduction of the SO interaction at large $\delta V_0$ is caused by the depletion in the proton $2s_{1/2}$ orbital between $^{36}$S and $^{34}$Si.

 Lowering the depth of the confining potential tends to shift the single-particle energies towards the continuum. As  found in \cite{Kay17}, the weak binding energy of certain orbitals then induces a departure from the linear behavior, displayed by dashed lines. At a binding energy of about -426 keV (-1.56 MeV), that is the experimental value of the 1/2$^-$ state  in $^{35}$Si ($^{37}$S), the calculated SO splitting amounts to about 0.93 (2 MeV).  The SO splitting value derived from the extrapolated linear trend is 1.03 MeV in $^{35}$Si. When compared to 0.93, it corresponds to a decrease of about 10\%. The calculated reduction in $2p$ SO splitting between $^{37}$Si  and $^{37}$S amounts to 1.07 MeV, among which 10\% is due to the effect of the continuum. 

\begin{figure*}[t]
\includegraphics[width=16cm]{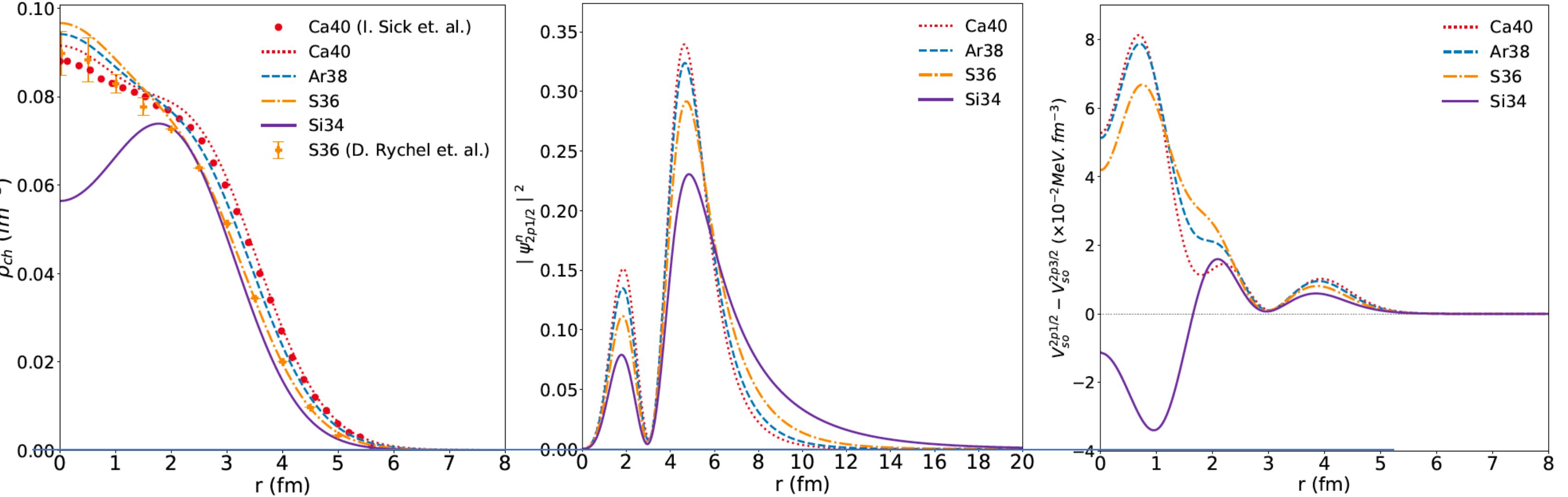}
\caption{Left: Calculated charge density distributions in the $N=20$ isotones. The experimental data points of Refs. \cite{Sick79,Rych83} for $^{40}$Ca and $^{36}$S are shown for comparison. Center: $2p_{1/2}$ squared radial wave functions in $^{41}$Ca, $^{39}$Ar, $^{37}$S and $^{35}$Si, in which the orbit is bound by 4.42, 3.34, 1.56 and 0.426 MeV, respectively. Right: Difference in SO potentials  felt by the $2p_{1/2}$ and $2p_{3/2}$ orbitals in the $N=20$ nuclei. All calculations are made with the DD-ME2 interaction. See text for details.}
\label{Dens} 
\end{figure*}

The calculated and, whenever available, experimental charge density distributions are displayed for the $N=20$ isotones in the left Fig. \ref{Dens}, respectively. Given the fact that the present DD-ME2 calculations do not include nuclear correlations, experimental and calculated charge density distributions of $^{40}$Ca and $^{36}$S are in rather good agreement. The calculated central depletion in $^{34}$Si is caused by the lack of protons in the $2s_{1/2}$ orbital.  The radial wave function of the weakly bound neutron $2p_{1/2}$ orbital  in $^{35}$Si, shown in the center of Fig. \ref{Dens}, extends to larger distances as compared to the other nuclei. A direct consequence is that the amplitude of the radial wave function between 4 and 6 fm, before the total and spin-orbit potentials drop to zero (see  Fig. \ref{pot}), is smaller than for the other $N=21$ isotones. The change in SO potential between the neutron $2p_{3/2}$ and $2p_{1/2}$ orbitals is displayed in the right part of Fig. \ref{Dens} for the $N=20$ isotones. Integrating this quantity over the nuclear volume gives rise to the calculated SO contribution to the $2p_{3/2}- 2p_{1/2}$ energy difference, $\Delta_{SO}$. In $^{35}$Si, the interior's contribution to the SO splitting has an opposite sign (meaning that the $2p_{1/2}$ would lie below the $2p_{3/2}$ orbital), while that at the surface has same sign as other isotones but is smaller.
 


As shown in Fig. \ref{Delta-SO}, the calculated evolution in SO splitting (blue filled squares) nicely follows, after a global offset correction of their amplitude, the experimental trend obtained from the energy of the major $2p_{3/2}$ and $2p_{1/2}$ fragments (red filled triangles). As opposed to the slight increasing trend of Ref. \cite{Mair93} (green stars), the calculated $p$ SO splitting is almost constant between $^{41}$Ca and $^{37}$S,  and suddenly drops  between $^{37}$S and $^{35}$Si, when protons are removed from the $2s_{1/2}$ orbital. 

{\it Other related cases - }Recent experiments have been carried out to determine the SO splittings of neutron and proton orbitals around $^{132}$Sn and $^{56}$Ni. As for the first region, Taprogge {\it et al.} \cite{Tapr14} determined the energy of the 3/2$^-$ proton state in $^{131}$In using the beta and beta-delayed neutron decays of $^{131}$Cd and $^{132}$Cd, respectively. A  $2p$ proton SO splitting of 988 keV was determined by combining this result with the previously known energy of the 1/2$^-$ isomeric state.  As shown in Fig. \ref{SO-Mairle}, the proton $2p$ SO splitting falls exactly on the global trend. 

Jones {\it et al.} \cite{Jone10} and Orlandi {\it et al.} \cite{Orla18}  independently studied the SO splittings for the valence neutron $2f$ and $3p$ and occupied states $2d$  around the doubly-magic $^{132}$Sn nucleus, respectively. Owing to the experimental difficulties to perform these studies and the fact that $^{132}$Sn is a doubly magic nucleus, they made the crude approximation that most of the single-particle and single-hole strengths were contained in one state. 

Jones {\it et al.} \cite{Jone10} used the $^{132}$Sn$(d,p)^{133}$Sn reaction to determine the $C^2S$ values of the 7/2$^-$, 3/2$^-$, 1/2$^-$ and 5/2$^-$ states, bound by about 2.4, 1.55, 1.04 and 0.4 MeV, respectively. Orlandi {\it et al.} \cite{Orla18} used the $^{132}$Sn$(d,t)^{131}$Sn reaction to determine the hole strength of the 3/2$^+$ ground state and 5/2$^+$ excited state at 1654 keV. The resulting SO splittings for the  $2f_{7/2}-2f_{5/2}$ (2 MeV),  $3p_{3/2}-3p_{1/2}$ (0.51 MeV), and $2d_{5/2}-2d_{3/2}$ (1.65 MeV) SO splittings are reported with diamond symbols in Fig. \ref{SO-Mairle}. Based on Woods-Saxon potential calculations with global parametrization, Ref. \cite{Orla18} pointed out that the SO splitting for the $2f$ and $3p$ orbitals deviate from systematics, while the $2d$ splitting has the expected value. These conclusions are qualitatively supported by the placement of the corresponding values, shown with diamond symbols, in Fig. \ref{SO-Mairle} :  the $3p$ case  lies about 24\% below the global trend, the $2f$ about 14\% below, and the $2d$ right on it. 

By using the same procedure as described for Fig. \ref{SO},  calculations performed with the DD-ME2 interaction predict a reduction of the $3p$ SO splitting by about 20\% due to the weak binding energy effect, in agreement with its placement below the trend of 
 Fig. \ref{SO-Mairle}. The combined low $\ell$ value and large number of nodes make the $3p$ wave function very sensitive to the effect of the continuum. Our predicted effect of the weak binding is however lower than the value of about 50\% estimated in Ref. \cite{Orla18}.  The present calculated effect of the continuum on the $2f$ SO splitting is negligible. Its value, slightly below the indicative trend of Mairle et al. \cite{Mair93}, could be explained  by some missing  $7/2^-$ hole or/and $5/2^-$ particle strengths around $^{132}$Sn, which would both shift the $2f$ SO splitting to larger values. This does not necessarily requires invoking an effect of the continuum, as in Ref. \cite{Orla18}. 


Finally Kahl  {\it et al.} have used the $^{56}$Ni$(d,p)^{57}$Ni and $^{56}$Ni$(d,n)^{57}$Cu reactions to determine $C^2S$ values of 0.73(31) and 0.66(22) for the 1/2$^-$ states  at 1.12 MeV and 1.109 MeV in $^{57}$Ni and $^{57}$Cu, respectively.  Within the error bars, these  $C^2S$ values almost exhaust the full single-particle strength. The spectroscopic factor values for the 3/2$^-$ ground states of  $^{57}$Ni and $^{57}$Cu could not be measured in their work as the experimental cross sections were deduced by  $\gamma$-ray  and not by  light particle detection.  However, assuming that the  $C^2S$ value of the g.s. was the same as that of the 1/2$^-$ state, a $2p$ SO splitting of about 1.1 MeV would be obtained for the two mirror nuclei. In spite of the uncertainties related to this assumption of similar $C^2S$ values, this $2p$ splitting is much lower than the one predicted by the trend of Fig. \ref{SO-Mairle}, which would rather lead to 1.65 MeV. It is also much lower than 2.02 MeV found in the $^{49}$Ca isotone using the major $2p$ fragments.  While the 1/2$^-$ state $^{57}$Cu is proton unbound by 420 keV and could be shifted to lower energy as compared to the mirror nucleus, there is no obvious reason for such a small SO splitting in $^{57}$Ni,  which is well bound and does not a priori exhibit a central density depletion. The DD-ME2 calculations found a reduction of the neutron $2p$ SO splitting  by only 10\% between $^{49}$Ca and $^{57}$Ni, in accordance with the trend of Mairle.


{\it Conclusions - } When using the self-consistent Covariant Energy Density Functional calculations with the DD-ME2 parametrization, it is found that the $2p_{3/2}- 2p_{1/2}$ SO splitting remains almost constant in the $N=21$ isotones between $^{41}$Ca ($Z=20$) and $^{37}$S ($Z=16$) and drops suddenly at $^{35}$Si ($Z=14$).  We propose that this effect is mainly caused by the depletion of the proton $2s_{1/2}$ orbital between $^{37}$S and $^{35}$Si, to which 10\% adds from the lingering of the weakly bound $2p$ orbits in $^{35}$Si. A similar conclusion has been derived from the shell-model + WS calculations of Ref. \cite{Otsu20}, in which the main reduction factor is caused by 2-body LS forces that induce an attractive (repulsive) interaction between the protons in the $2s_{1/2}$  and  the neutrons in the $2p_{3/2}$ ($2p_{1/2}$) orbitals, to which an additional 8\% is due to the proximity of the continuum. 

We  have also shown in the present work that adding a central density depletion to a simple WS potential induces a further SO reduction, as compared to the work of Kay {\it et al.} \cite{Kay17}, in which only the effect of weak binding was considered. Therefore, neglecting the central density depletion leads to an overestimation of the weak binding energy effect on the SO reduction. 

Our conclusions are at variance with those of Kay {\it et al.} which proposed that "the proximity of the threshold can explain the (full) reduction of the splitting between these (neutron $2p_{3/2}- 2p_{1/2}$) spin-orbit partners" and that "this effect must be taken into account {\it before} other explanations are considered". We have demonstrated in the present work, complementary Refs. \cite{Sign07, Gaud07, Burg14, Li16, Dugu17, Otsu20}, that all effects, including those induced by realistic nuclear forces, correlations, and the proximity of the continuum, must be taken into account on the same footing to reach a complete understanding of the SO splitting in the $N=21$ isotones.

Applying our self-consistent calculations to the valence states in $^{132}$Sn confirms the conclusion reached in Ref. \cite{Orla18} that  the continuum plays a significant role in reducing the neutron $3p$ SO splitting (though the reduction we calculate is weaker), but we find find no reduction of any great size in the $2f$ SO splitting. The significant reduction of the neutron $2p$ SO splitting in $^{57}$Ni, compared to $^{49}$Ca, remains without explanation. It would be worthwhile for further experimental studies to be carried out in $^{57}$Ni in order to clarify the situation. Such data would certainly benchmark the $A \simeq 60$ region that so far  has been poorly studied, but where the most significant deviations are observed relative to the global mass dependence, as shown in Fig. \ref{SO-Mairle}.



\section{Acknowledgments}
O. S wishes to thank T. Duguet, F. Nowacki, A. Macchiavelli and M. Ploszajczak for fruitful discussions and encouragements to publish this work.



\begin{thebibliography}{100}

\bibitem{Ebra16} J.-P. Ebran {\it et al.}, J. Phys. G {\bf 43} (2016) 085101.
\bibitem{Gopp49} M. G. Mayer Phys. Rev. C {\bf 75} (1949) 1969; O. Haxel, J. H. D Jensen and H. E. Suess, Phys. Rev. C {\bf 75} (1949) 1766.
\bibitem{Mair93} G. Mairle, Phys. Lett. B {\bf 304} (1993) 39.
\bibitem{Sorl08} O. Sorlin and M.-G. Porquet, Prog. in Part. and Nucl. Phys. {\bf 61} (2008) 602.
\bibitem{Gaud06} L. Gaudefroy {\it et al.} Phys. Rev. Lett. {\bf 97} (2006) 092501. 

\bibitem{Smir12} N. A. Smirnova, K. Heyde, B. Bally, F. Nowacki, and K. Sieja, Phys. Rev. C {\bf 86}, 034314 (2012).
\bibitem{Otsu20} T. Otsuka {\it et al.}, Rev of Mod. Phys. \textbf{92}  (2020) 015002.
\bibitem{Bend99} M. Bender {\it et al.}, Phys. Rev. C {\bf 60} (1999) 034304.
\bibitem{Shar95} M. M. Sharma  {\it et al.}, Phys. Rev. Lett. {\bf 72} (1994) 1431.
\bibitem{Lala98} G. A. Lalazissis {\it et al.}, Phys. Lett. B {\bf 418}  (1998) 7.
\bibitem{Ebra16b} J.-P. Ebran {\it et al.}, Phys. Rev. C {\bf 94}  (2016) 024304.
\bibitem{Todd04} B. G. Todd-Rutel, J. Piekarewicz, and P. D. Cottle, Phys. Rev. C \textbf{69} (2004) 021301(R) 
\bibitem{Li16} J. J. Li {\it et al.}, Phys. Rev. C {\bf 93}  (2016) 054312.
\bibitem{Muts17} A. Mutschler {\it et al.}, Nature Physics {\bf 13} (2017) 152. 
\bibitem{Dugu17} T. Duguet {\it et al.} Phys. Rev. C {\bf 95} (2017) 034319.
\bibitem{Grass09} M. Grasso  {\it et al.},  Phys. Rev. C  {\bf 79}  (2009) 034318.



\bibitem{Sign11} A. Signoracci, B. A. Brown and M. Hjorth-Jensen, Phys. Rev. C {\bf 83}  (2011) 024315.
\bibitem{Yuan14} C. Yuan {\it et al.}, Phys. Rev. C {\bf 89} (2014) 044327. 
\bibitem{Mich05} N. Michel {\it et al.}, Nucl. Phys. A  {\bf 752} (2005) 335c.

\bibitem{Cave82} J. M. Cavedon \emph{et al.}, Phys. Rev. Lett. \textbf{49} (1982) 978.
\bibitem{Saxe19} G. Saxena {\it et al.}, Phys. Lett. B \textbf{788}  (2019) 1. 


\bibitem{Burg14} G. Burgunder {\it et al.}, Phys. Rev. Lett. {\bf 112} (2014) 042502. 

\bibitem{Kay17} B. P. Kay, C. R. Hoffman and A. O. Macchiavelli, Phys. Rev. Lett. {\bf 119}  (2017) 182502.
\bibitem{Jone10} K. L. Jones {\it et al.}, Nature {\bf 465}  (2010) 454. 
\bibitem{Tapr14} J. Taprogge {\it et al.}, Phys. Rev. Lett. {\bf 112} (2014) 132501. 
\bibitem{Orla18} R. Orlandi {\it et al.}, Phys. Lett. B \textbf{785}  (2018) 615.
\bibitem{ENSDF}  From ENSDF database, https://www.nndc.bnl.gov/ensdf 
\bibitem{Dugu15} T. Duguet {\it et al.}, Phys. Rev. C {\bf 92} (2015) 034313. 




\bibitem{Hamm01} I. Hammamoto, Nucl. Phys. A {\bf 683} (2001) 255.

\bibitem{Fort95} H. T. Fortune, D  Koltenuk and C. K.  Lau, Phys. Rev. C {\bf 51} (1995) 3023. 
\bibitem{Stef14} I. Stefan  {\it et al.},  Phys. Rev. C  {\bf 90}  (2014) 014307.



\bibitem{DDME2} G. A. Lalazissis {\it et al.}, Phys. Rev. C {\bf 71} (2005) 024312. 
\bibitem{ber03} J.F. Berger, D. Hirata and M. Girod, Act. Phys. Pol. B \textbf{34}  (2003) 1909. 
\bibitem{per08} S. Perez-Martin and L. M. Robledo, Phys. Rev. C \textbf{78}  (2008) 014304. 
\bibitem{koo34} T. Koopmans, Physica \textbf{1} (1934) 104. 
\bibitem{Sick79} I. Sick {\it et al.}, Phys. Lett. B \textbf{88}  (1979) 245.
\bibitem{Rych83} D. Rychel  {\it et al.}, Phys. Lett. B \textbf{130}  (1983) 5.

\bibitem{Kahl19} D. Kahl  {\it et al.}, Phys. Lett. B \textbf{797}  (2019) 134803.





\bibitem{Sign07} A. Signoracci and B. A. Brown, Phys. Rev. Lett.  {\bf 99} (2007) 099201. 


\bibitem{Gaud07} L. Gaudefroy {\it et al.}, Phys. Rev. Lett. {\bf 99} (2007) 099202. 


\end{thebibliography}
\end{document}